\documentclass[aps,showpacs,prl,reprint,superscriptaddress]{revtex4-1}

\usepackage[colorlinks=true,linkcolor=blue,citecolor=blue,urlcolor=blue]{hyperref}
\usepackage{setspace} 
\usepackage{graphicx}
\usepackage{amsmath}
\usepackage{color}
\usepackage{amsmath}
\usepackage{amssymb}
\usepackage{verbatim}
\usepackage{latexsym}
\usepackage{enumerate} 
\usepackage{bm} 

\begin{document}

\title{Electron-Phonon Coupling and the Metalization of Solid Helium at Terapascal Pressures} 


\author{Bartomeu Monserrat}
\email{bm418@cam.ac.uk}
\affiliation{TCM Group, Cavendish Laboratory, University of Cambridge,
  J.\ J.\ Thomson Avenue, Cambridge CB3 0HE, United Kingdom}

\author{N.\ D.\ Drummond}

\affiliation{Department of Physics, Lancaster University, Lancaster
  LA1 4YB, United Kingdom}

\author{Chris J.\ Pickard}

\affiliation{Department of Physics and Astronomy, University College London, Gower Street, London WC1E 6BT, United Kingdom}

\author{R.\ J.\ Needs}

\affiliation{TCM Group, Cavendish Laboratory, University of Cambridge,
  J.\ J.\ Thomson Avenue, Cambridge CB3 0HE, United Kingdom}

\date{\today}

\begin{abstract}
  Solid He is studied in the pressure and temperature ranges
  1--40 TPa and 0--10,000 K using first-principles methods.
  Anharmonic vibrational properties are calculated within a
  self-consistent field framework, including the internal
  and free energies, density-pressure relation, stress tensor, thermal
  expansion, and the electron-phonon coupling renormalization of the
  electronic band gap.
  We find that an accurate description of electron-phonon coupling
  requires us to use a non-perturbative approach.
  The metalization pressure of $32.9$ TPa at $0$ K\@ is larger
  than found previously.
  The vibrational effects are large; for example at $P=30$~TPa the band gap is
  increased by $2.8$ eV by electron-phonon coupling and a further $0.1$ eV
  by thermal expansion compared to the static value.
  The implications of the calculated metalization pressure for the
  cooling of white dwarfs are discussed.
\end{abstract}

\pacs{63.20.Ry, 71.38.-k, 62.50.−p}

\maketitle




Helium (He) is the second most abundant element in the Universe after
hydrogen and one of the most important components of
stellar bodies such as giant gaseous planets, main-sequence stars, and
white dwarf (WD) stars.
The large value of the first excitation energy of atomic He of $19.82$ eV
leads to a high metalization pressure for the solid phase, of the
order of tens of terapascals. Calculations of the phase diagram of
He \cite{PhysRevB.24.5119,LPB:4324892} indicate that, in the
terapascal pressure range, it remains solid up to
temperatures of around $8,000$ K\@.  He is therefore
expected to be found in the solid state in the outer layers of cool
WDs.

The vast majority of the stars in the Universe become WDs in the final
stages of their evolution, with the gravitational attraction towards
the center being balanced by the electron degeneracy pressure of the
high-density core. The lack of a continuous energy source means that
WDs cool down until reaching thermodynamic equilibrium with their
surroundings, eventually becoming black dwarfs. An understanding of
the cooling process \cite{annurev.aa.28.090190.001035} is essential when
calculating the ages of observed WDs, which are widely used within
cosmochronology \cite{1987ApJ315L77W,cosmocrhonology} to date stellar
clusters and galaxies, and hence to provide bounds on the age of the
Universe.

The cores of WDs are largely isothermal due to the high
thermal conductivity of degenerate electrons. Hence the cooling rate
is mainly determined by the outer layers, which are composed of
hydrogen, He, or a mixture of both. In this context the
insulator-metal transition in solid He is central because
energy transport from the degenerate core is dominated by electron
transport through metallic He in the deeper layers, and by photon
transport through insulating He in the outermost
region \cite{PhysRevLett.101.106407}.

Recent work has focused on the study of the metalization pressure of He in
the solid and fluid states
\cite{PhysRevLett.101.106407,Stixrude12082008,PhysRevB.76.075112,PhysRevLett.104.184503,PhysRevB.86.115102,ceperley_rmp_h_he}.
In the solid state \cite{PhysRevLett.101.106407}, static-lattice electronic
structure calculations using both the diffusion Monte Carlo (DMC) many-body
wave function technique \cite{PhysRevLett.45.566,RevModPhys.73.33} and the
\textit{GW} approximation of many-body perturbation theory
\cite{0034-4885-61-3-002} have shown that standard generalized gradient
approximations \cite{PhysRevLett.77.3865} to density functional theory (DFT)
\cite{PhysRev.136.B864,PhysRev.140.A1133} substantially underestimate band
gaps. In the fluid state \cite{Stixrude12082008}, the effects of
electron-phonon coupling lead to a strong temperature dependence of the
metalization pressure. The effects of temperature on the metalization transition in the solid state remain an open question.



He is the second lightest element and therefore the amplitudes of the
nuclear vibrations are expected to be large and possibly to present anharmonic behavior. Recent advances in the treatment of anharmonicity from first-principles make the incorporation of these effects feasible \cite{PhysRevLett.100.095901,PhysRevLett.106.165501,PhysRevB.84.180301,PhysRevB.86.054119,PhysRevB.87.144302,errea_arxiv}.
For the determination of band gaps in the solid state, both electron-phonon coupling and thermal expansion are important \cite{RevModPhys.77.1173,0022-3719-9-12-013,PhysRevB.23.1495}. Studying the
effects of electron-phonon coupling on the band gaps of solids from first
principles has only recently become possible
\cite{0295-5075-10-6-011,PhysRevB.73.245202,PhysRevLett.105.265501,PhysRevB.87.144302,giustino_nat_comm}. In this paper, we combine the first-principles calculation of anharmonicity, electron-phonon coupling and thermal expansion
to study the vibrational corrections to the
thermal band gap of solid He and hence determine an accurate value for the
metalization pressure including the effects of temperature.






We use the principal-axes approximation \cite{PhysRevB.87.144302,jung:10332} to the
Born-Oppenheimer energy surface, 
considering independent phonon terms and pairwise phonon-phonon interactions. 
The resulting Schr\"{o}dinger equation for the nuclear coordinates is solved
within the vibrational self-consistent field (VSCF) framework
\cite{PhysRevB.87.144302,bowman:608} for the vibrational anharmonic energy
$E_{\mathbf{S}}$ and wave function $|\Phi^{\mathbf{S}}(\mathbf{Q})\rangle$ in
state $\mathbf{S}$, where $\mathbf{Q}$ is a collective phonon coordinate. We use second-order perturbation theory to go beyond the
mean-field formulation.
We then calculate phonon expectation values of a general operator $\hat{O}(\mathbf{Q})$ 
at zero and
finite temperature $T$ according to
\begin{equation}
  \langle \hat{O}(\mathbf{Q})\rangle=\frac{1}{\mathcal{Z}}\sum_{\mathbf{S}}\langle\Phi^{\mathbf{S}}(\mathbf{Q})|\hat{O}(\mathbf{Q})|\Phi^{\mathbf{S}}(\mathbf{Q})\rangle e^{-E_{\mathbf{S}}/k_{\mathrm{B}}T}, \label{eq:ph_exp}
\end{equation}
where $\mathcal{Z}$ is the partition function and $k_{\mathrm{B}}$ is
Boltzmann's constant. This expectation value can be calculated by writing the operator $\hat{O}(\mathbf{Q})$ within a principal axes representation like that for the energy \cite{PhysRevB.87.144302}, or non-perturbatively by sampling phase space according to the nuclear density \cite{PhysRevB.73.245202,giustino_nat_comm}. The first method is approximate, but the description is in terms of individual phonons, permitting direct access to the underlying physical processes. The second method can in principle lead to accurate numerical results, but the underlying physical mechanisms are obscured. Here we will use a combination of both methods to obtain a full picture of the effects of electron-phonon coupling on the band gap of solid He. We note that 
the use of DFT to calculate electron-phonon corrections
to band gaps is reasonable, as the usual band-gap underestimation cancels in the difference between the static gap and the phonon-renormalized gap \cite{PhysRevLett.105.265501}.


\begin{figure}
\centering
\includegraphics[scale=0.32]{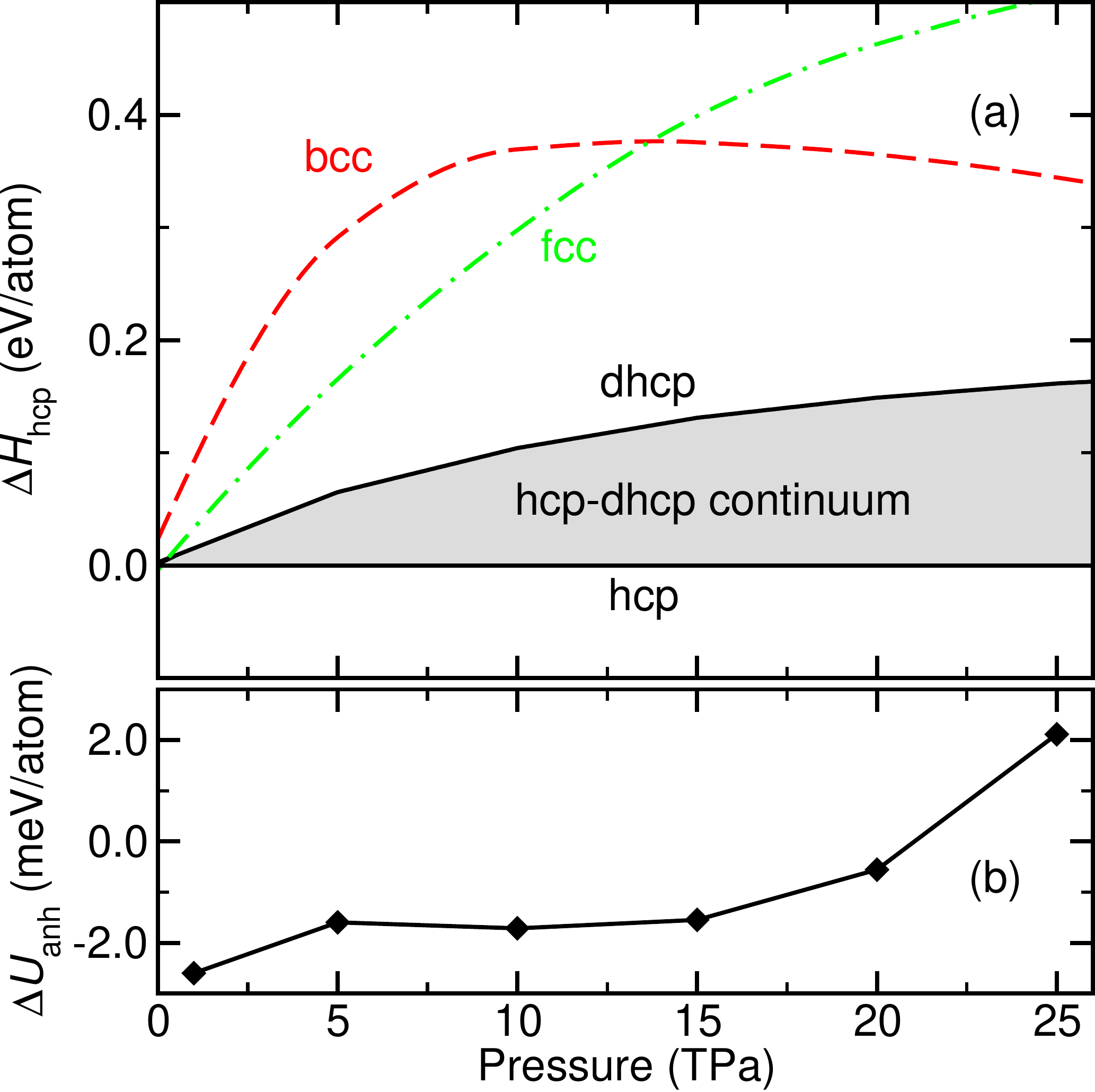}
\caption{(color online) (a) Enthalpy $\Delta H_{\mathrm{hcp}}$ with respect to the hcp phase of the closed-packed phases described at the static DFT level. (b) Anharmonic energy correction $\Delta U_{\mathrm{anh}}$ to the
  harmonic energy of the hcp phase at zero temperature.}
\label{fig:phases}
\end{figure}


We have solved the electronic Schr\"{o}dinger equation 
to map the
Born-Oppenheimer energy surface within plane-wave DFT
\cite{PhysRev.136.B864,PhysRev.140.A1133} as implemented in the
\textsc{castep} code \cite{CASTEP}. We have used ultrasoft pseudopotentials
\cite{PhysRevB.41.7892} with core radii of $0.212$ {\AA}, which require an energy
cutoff of $2800$ eV, and Monkhorst-Pack \cite{PhysRevB.13.5188} ${\bf
  k}$-point grids of spacing $2\pi\times0.04$ \AA$^{-1}$.  These parameters
lead to energy differences converged to within $10^{-4}$ eV per atom
and stresses to within $10^{-1}$ GPa. All calculations were performed with the
Perdew-Burke-Ernzerhof \cite{PhysRevLett.77.3865} generalized gradient
approximation density functional.
We solved the vibrational Schr\"{o}dinger equation within the
VSCF formalism by expanding the wave function in a basis of simple harmonic
oscillator (SHO) eigenstates. This basis is defined from a quadratic fit to the Born-Oppenheimer energy surface calculated within the principal axes approximation rather than from the harmonic
approximation. We have included $100$ SHO states for each phonon degree of freedom, which leads to converged results.


\begin{figure}
\centering
\includegraphics[scale=0.33]{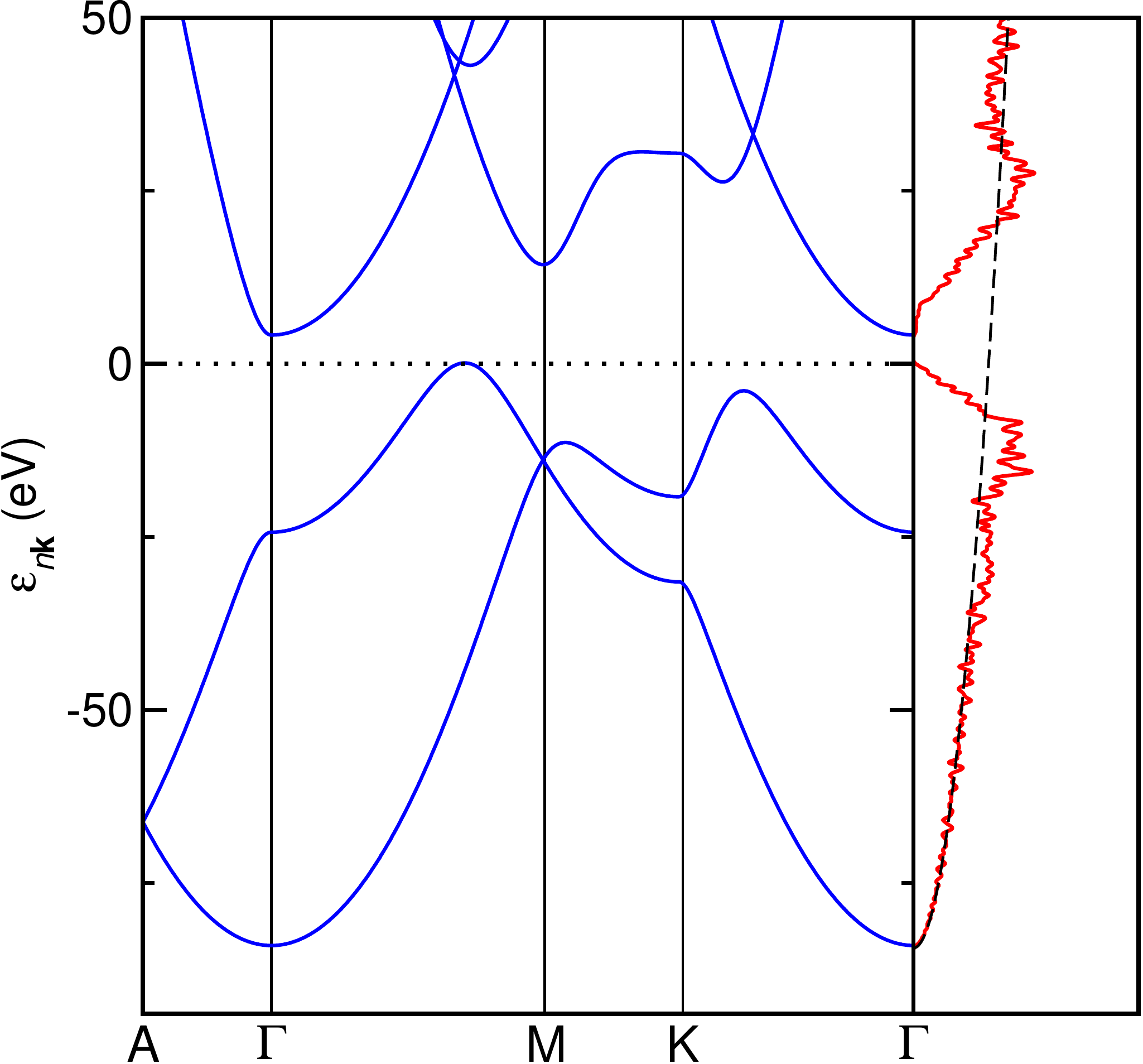}
\caption{(color online) Electronic band structure (blue) and density of states
  (red, on right-hand-side of figure) of hcp He at $10$ TPa. The dashed
  black line is the density of states of a free electron gas of the same
  density as He.}
\label{fig:bs}
\end{figure}

The phase diagram of solid He at low pressures is
well-characterized, and the hexagonal close-packed (hcp) structure
with a $c/a$ ratio close to ideal 
has been found to be stable up to at
least $58$ GPa \cite{PhysRevLett.71.2272}. It is usually assumed that
He remains in the solid hcp phase up to high pressures and
temperatures, but this does not seem to have been tested in detail. Furthermore, at the highest pressures it is expected to be in a body-centered cubic (bcc) phase.
We have therefore performed searches at $10$ and $20$ TPa using the \textit{ab initio} random structure searching (AIRSS)
method \cite{PhysRevLett.97.045504,Pickard2011} to find low-enthalpy crystal structures of He with $12$ or fewer atoms per cell. We have calculated the harmonic vibrational free energy of the most competitive phases in the temperature range 0--10,000 K, and several phases were found that
were lower in free energy than the face-centered cubic (fcc) and the bcc phases, but none of them were more
stable than hcp. In Fig.\ \ref{fig:phases} we show the static lattice enthalpy $\Delta H_{\mathrm{hcp}}$ with respect to the hcp enthalpy of the closed-packed phases as a function of pressure. The difference in enthalpy between these closed-packed phases varies with pressure, which shows that the electronic structures of the phases, as well as packing considerations, are important in determining the relative enthalpies. 
The hcp and double hexagonal closed-packed (dhcp) structures are the
two end members of a series of structures which differ in the stacking
of layers.  The energies per layer of these structures vary
continuously with the fraction of stacking faults present.


\begin{figure}
\centering
\includegraphics[scale=0.32]{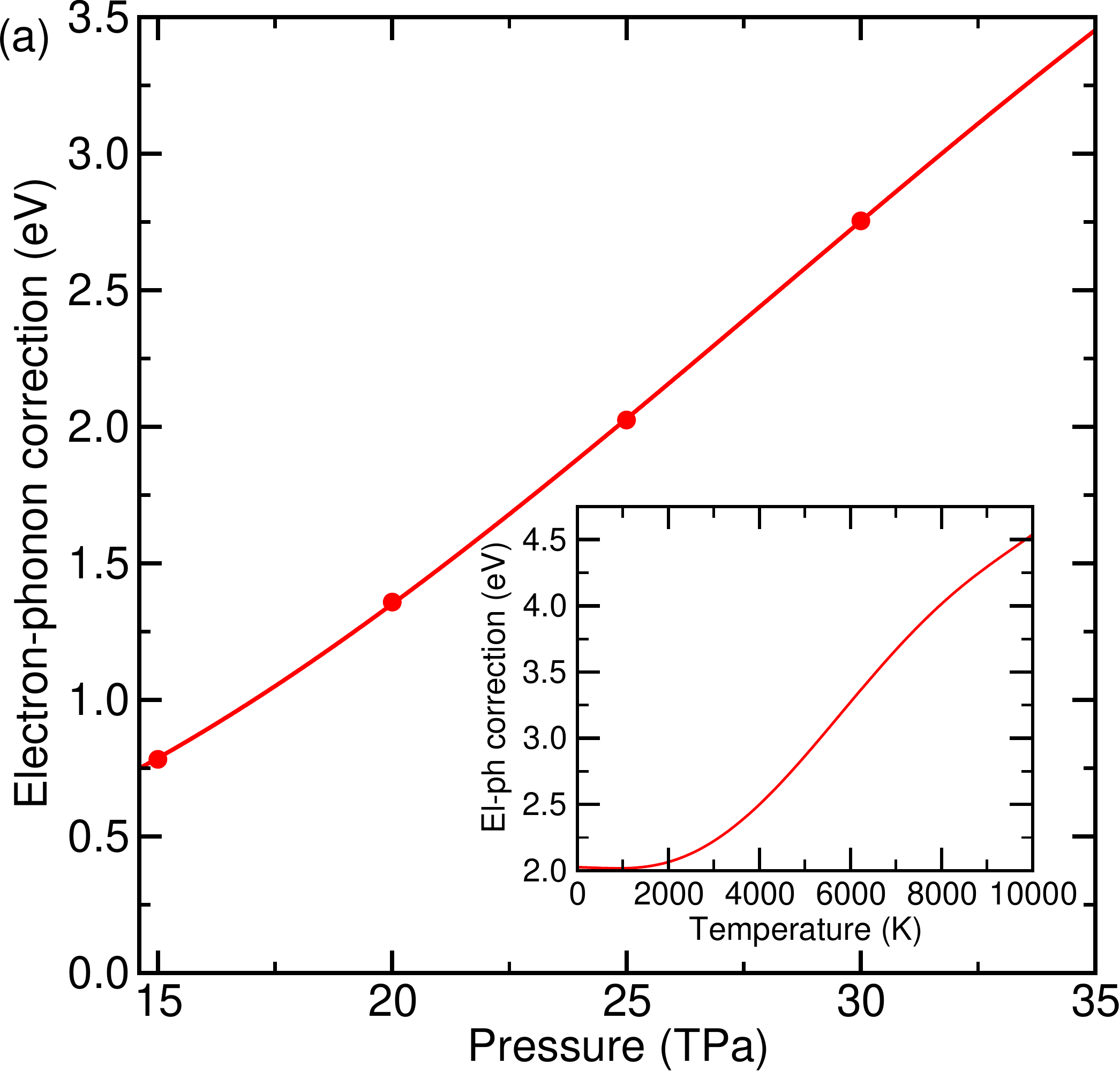}
\includegraphics[scale=0.38]{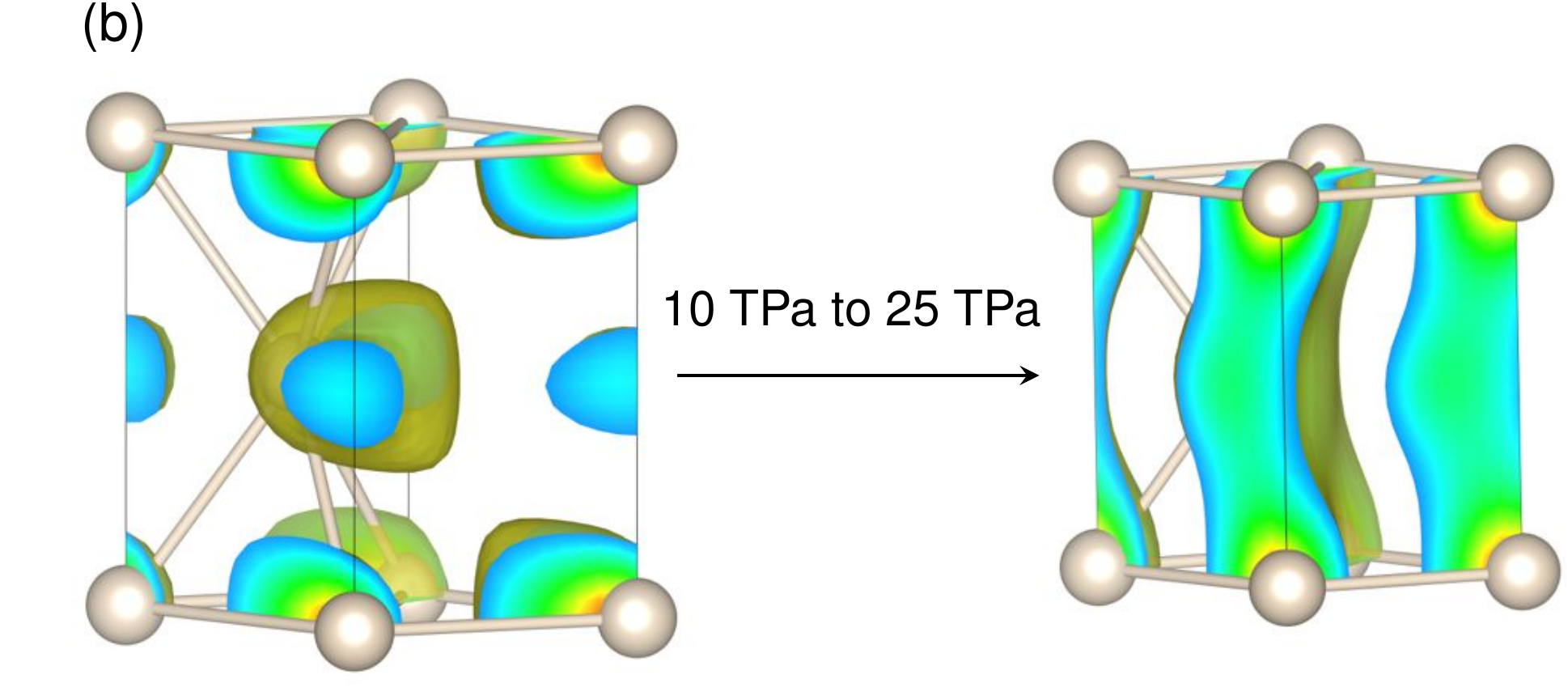}
\caption{(color online) (a) ZP electron-phonon correction to the thermal
  band gap of hcp solid He as a function of pressure. The inset
  shows the temperature dependence of the electron-phonon correction at a
  pressure of $P=25$ TPa. (b) Squared Kohn-Sham eigenstate corresponding to the CBM at $10$ TPa and $25$ TPa, using a R-G-B color scale with red corresponding to high density and blue to low density. We have removed the regions of lowest density for clarity.}
\label{fig:elph}
\end{figure}

We have also calculated the anharmonic free energy contribution $\Delta
F_{\mathrm{anh}}$ for the hcp structure (see Fig.\
\ref{fig:phases}). For example, at $10$ TPa and zero temperature the harmonic energy is $686.9$ meV/atom and the anharmonic correction is $-1.7$ meV/atom due to the independent phonon term, and it is further renormalized by $+1.1$ meV/atom due to the two-body phonon term. At $T=5,000$ K, the harmonic free energy is $75.0$ meV/atom, and the anharmonic correction is $-5.2$ meV/atom. At $25$ TPa and zero temperature the harmonic energy is $903.0$ meV/atom and the anharmonic correction is only $2.1$ meV/atom, whereas at $T=5,000$ K, the harmonic free energy is $466.6$ meV/atom and the anharmonic correction is $6.3$ meV/atom.
Second-order perturbation theory does not change these results within the reported precision, demonstrating the accuracy of the mean-field approximation. These anharmonic corrections, albeit larger than in heavier systems such as diamond \cite{PhysRevB.87.144302}, remain remarkably small and have no discernible effect on the relative stability of the phases considered. Furthermore, the ratio of the anharmonic to quasiharmonic vibrational free energy decreases with increasing pressure, 
which suggests that anharmonicity becomes less important for the energetics of He at high pressures.

\begin{figure}
\centering
\includegraphics[scale=0.32]{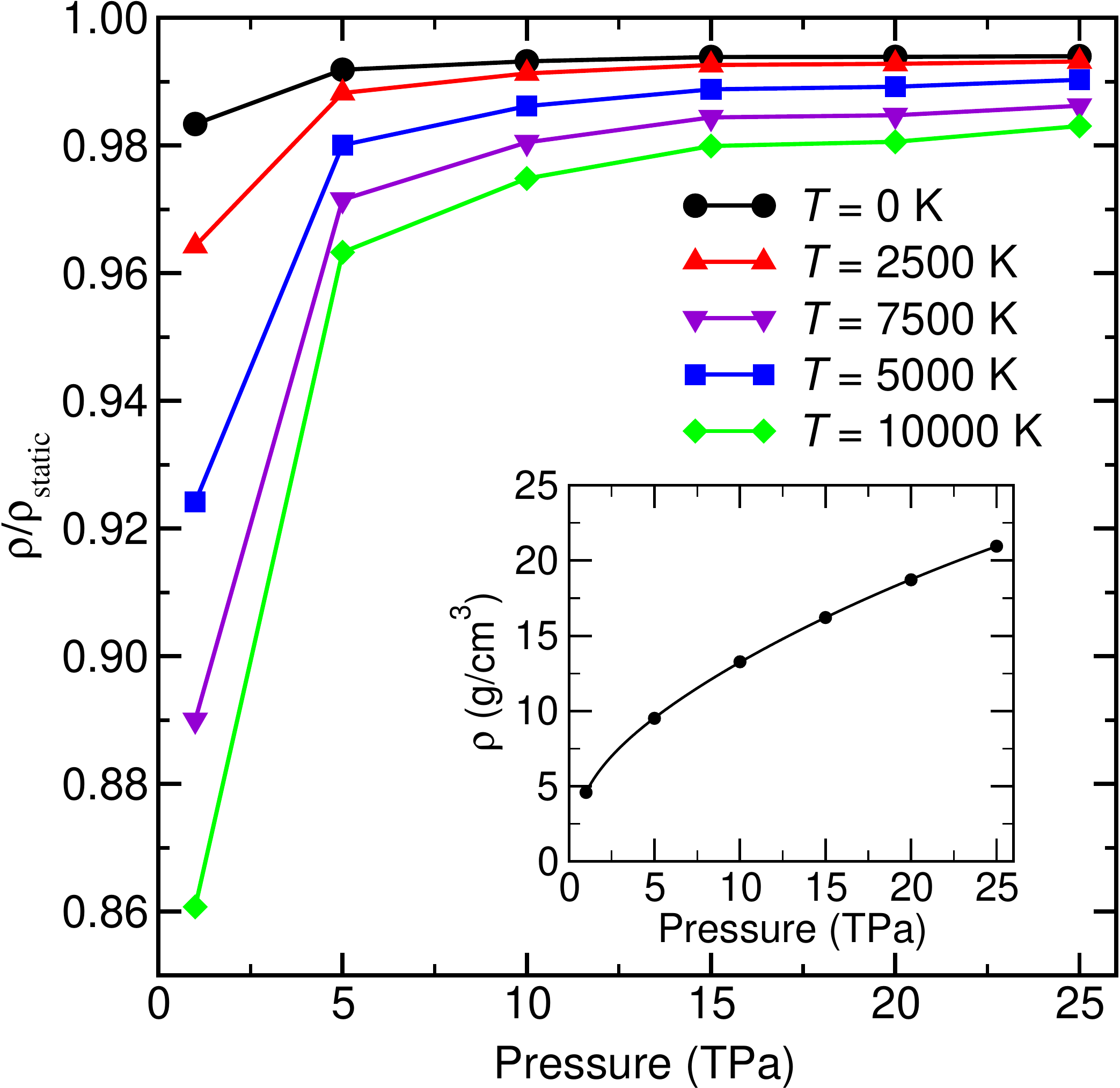}
\caption{(color online) Density-pressure-temperature phase diagram of
  solid hcp He. At a given external pressure, the density $\rho$
  is renormalized with respect to the static-lattice density
  $\rho_{\mathrm{static}}$ at that pressure. The inset shows the
  density-pressure relation at $T=0$ K, including the effects of
  ZP motion.}
\label{fig:density}
\end{figure}


The band structure and density of states of solid hcp He at $10$ TPa are
shown in Fig.\ \ref{fig:bs}. The density of states of the occupied bands is
very close to that of a free electron gas at the same
density, reflecting the free-electron nature of high-pressure He. The valence band maximum (VBM)
is located along the symmetry line joining the $\Gamma$ and $M$ points, while
the conduction band minimum (CBM) is located at the $\Gamma$ point, as observed experimentally at lower pressures \cite{PhysRevLett.105.186404}. We have evaluated the expression in Eq.\ (\ref{eq:ph_exp}), with $\hat{O}$  becoming the difference between the CBM and the VBM,  both within a principal axes representation and non-perturbatively 
to investigate the band-gap renormalization
of solid hcp He due to electron-phonon coupling \footnote{Supercells containing $54$ atoms are used for all reported calculations. Tests with supercells containing up to $250$ atoms show a convergence of the band gap corrections within $0.2$ eV, leading to an uncertainty of $0.4$ TPa in the metallization pressure. For the non-perturbative calculations, a total of $1,000$ sampling points have been used at each pressure and temperature, leading to statistical error bars smaller than the reported accuracy}. 
Figure\ \ref{fig:elph} shows the pressure dependence of the
zero-point (ZP) electron-phonon correction to the thermal band gap of hcp solid He calculated non-perturbatively.
A second calculation of the same quantity with a principal axes representation of $E_{\mathrm{g}}(\mathbf{Q})$ gives information about the underlying physical mechanisms.
The out-of-plane modes of the hcp lattice couple strongly to the electronic bands and their effect is to open the band gap. This coupling increases with pressure, and it leads to the behavior shown in Fig.\ \ref{fig:elph}. This can be further understood by analyzing the Kohn-Sham eigenstates corresponding to the VBM and the CBM (shown in Fig.\ \ref{fig:elph}). At lower pressure these eigenstates are localized around the atomic sites, but as pressure increases they delocalize in the interplane direction, hence increasing the coupling with the out-of-plane modes. 
We emphasize that the non-perturbative calculation of the renormalization of the band gap due to electron-phonon coupling leads to quantitatively different results to the perturbative approach. Terms beyond lowest-order perturbation theory are found to be crucial for calculating the correction to the band gap \cite{analytic_arxiv}. The magnitude of the band gap renormalization is, as far as we are aware, amongst the largest reported, similar only to that found in high-pressure solid hydrogen~\cite{ceperley_h_elph_coupling}.

\begin{figure}
\centering
\includegraphics[scale=0.32]{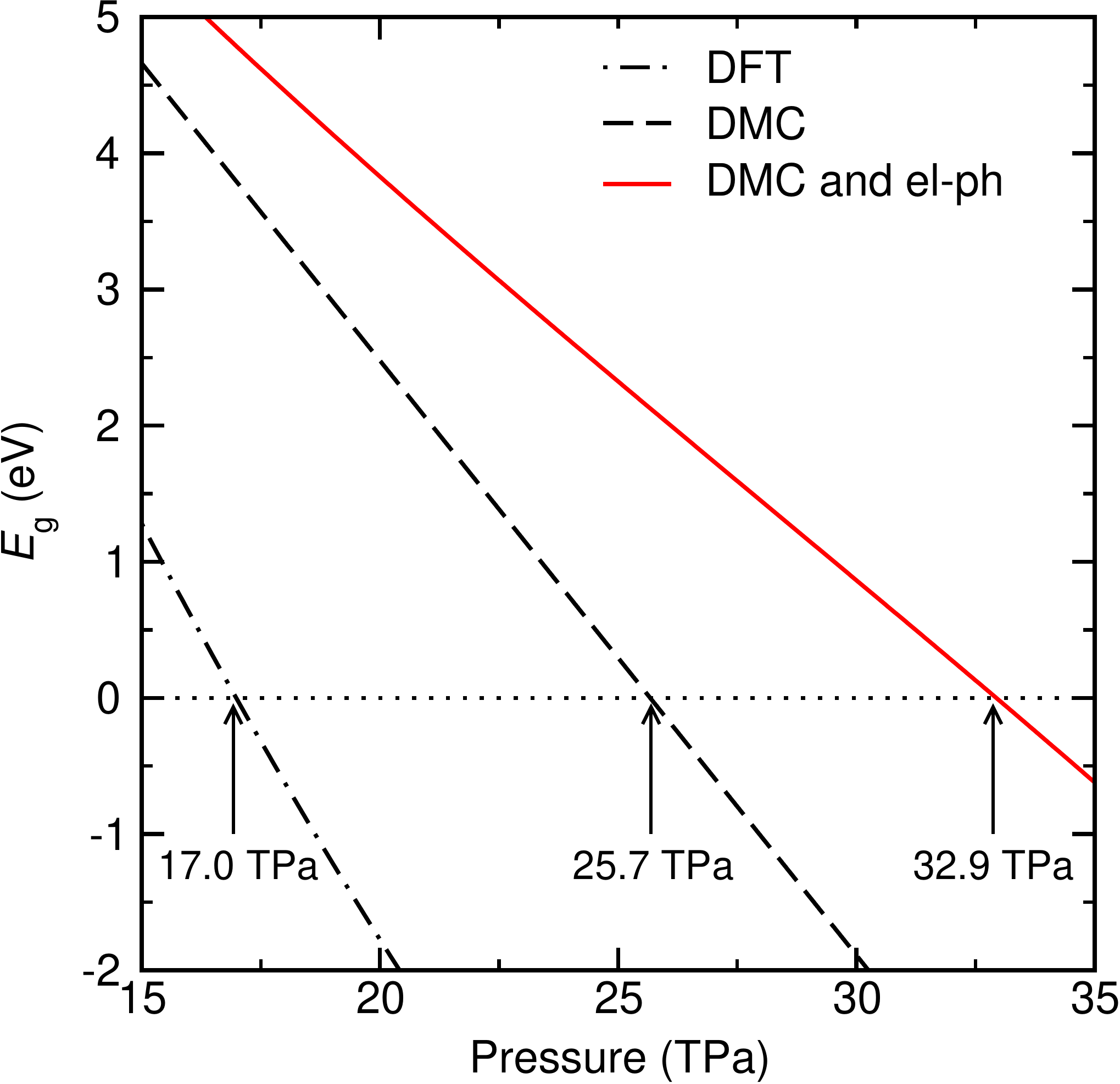}
\caption{(color online) Thermal band gap $E_{\mathrm{g}}$ of solid hcp He
  as a function of pressure, calculated by evaluating the difference between the CBM and the VBM (even when the difference is negative). The static gap results correspond to DFT calculations (dashed-dotted black line) and DMC calculations (dashed black line)
  taken from
  Ref.\ \cite{PhysRevLett.101.106407}. The band gap including the effects
  of electron-phonon coupling and thermal expansion added to the DMC results is shown at $T=0$ K (red
  solid line).} 
\label{fig:metal}
\end{figure}

To investigate thermal expansion we have calculated the vibrational stress tensor at finite temperatures using the formalism described in Ref.\ \cite{PhysRevB.87.144302}. The vibrational stress tensor is diagonal, with similar
in-plane and out-of-plane stresses, leading to an isotropic volume expansion. Figure\ \ref{fig:density} shows the density-pressure relation at five
different temperatures, including the full vibrational effects. Both
ZP and finite-temperature expansions are more significant at
lower pressures. This occurs because the system stiffens with
increasing pressure (see the inset of Fig.\ \ref{fig:density}), and 
although the vibrational stress tensor is larger at higher pressures, the resulting density change is smaller.
At $P=25$ TPa, the ZP correction to the volume at $T=0$ K
opens the gap by $0.105$ eV, and the thermal expansion at
$T=5,000$ K opens the gap by $0.216$ eV\@.


Combining the effects of electron-phonon coupling and thermal expansion on the band gap, we calculate the pressure dependence of the thermal band gap of solid He near metalization, as shown in Fig.\ \ref{fig:metal}. The static-lattice DFT value has been calculated using the PBE functional. The static-lattice DMC 
results are those reported in Ref.\ \cite{PhysRevLett.101.106407}, where \textit{GW} calculations lead to results matching those from DMC calculations.
The curve including electron-phonon coupling and ZP expansion at zero temperature leads to a metalization pressure of $32.9$~TPa, significantly higher than the static-lattice value. 

\begin{figure}
\centering
\includegraphics[scale=0.73]{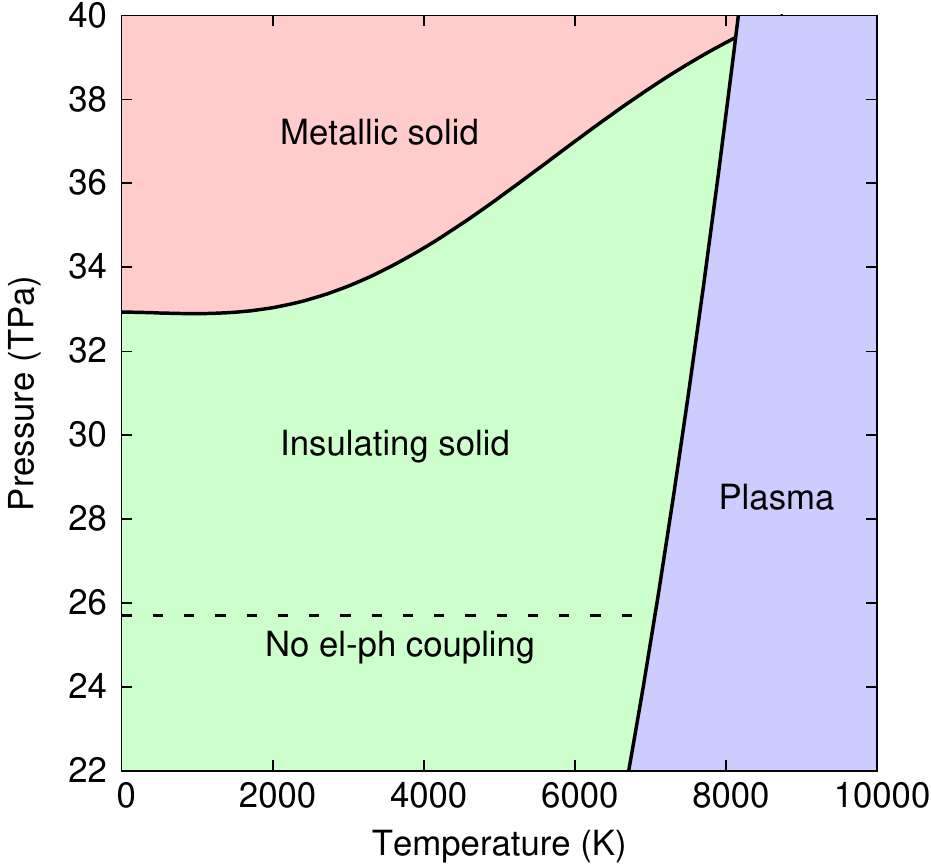}
\caption{(color online) Phase diagram of He at the pressure-temperature
  range relevant for cool WDs. The black dashed line is the insulator-metal
  transition predicted in Ref.\ \cite{PhysRevLett.101.106407}. The
  solid-plasma transition is taken from Ref.\ \cite{LPB:4324892}.}
\label{fig:phase_diagram}
\end{figure}

The renormalization of the band gap due to lattice vibrations near the
metalization pressure is in the range $3.0$--$4.0$ eV for temperatures
in the range $0$--$5,000$ K\@. These values represent a quantitative
change to the static lattice value of the same order of magnitude as the electron-correlation
correction to DFT \cite{PhysRevLett.101.106407}.
Furthermore, our results are the first to describe the temperature-dependence of the transition. The authors of Ref.\ \cite{PhysRevLett.101.106407} considered the effects of temperature on the electronic band gap within path integral Monte Carlo \cite{RevModPhys.67.279}, and concluded that they are negligible. We have come to a significantly different conclusion on this issue, but there is not enough information in Ref.\ \cite{PhysRevLett.101.106407} to be able to make a more detailed comparison.  From the considerations above and the data of Ref.\ \cite{LPB:4324892}, we can construct the phase diagram of He at the pressures and temperatures relevant for cool WDs as shown in Fig.\ \ref{fig:phase_diagram}.



In conclusion, we have presented first-principles calculations of the phase
stability, electron-phonon coupling, and thermal expansion in solid He for
a range of pressures and temperatures. We have shown that the thermodynamically stable phase of solid He is hcp for the pressure range
1--30 TPa and temperature range 0--10,000 K\@, including anharmonic energies at the mean-field level. The second-order perturbation theory used to go beyond the VSCF approximation gives a negligible correction, suggesting that
the mean field energy is accurate.  
The effects of electron-phonon coupling on the band gap are
substantial and several times larger than the effects of thermal
expansion.  A perturbative approach is not accurate here, and it is important to
use a non-perturbative scheme, as we have done.
We have determined the metal-insulator transition of solid He to
be at $32.9$ TPa at $T=0$ K\@, which may be compared with the
value $25.7$ TPa obtained in Ref.\ \cite{PhysRevLett.101.106407} at the static-lattice level.  

The increase in the metalization pressure of solid He when
the effects of electron-phonon coupling and thermal
expansion are included implies that the interiors of WDs have a metallic He
layer that is thinner than the one predicted by a purely electronic
treatment, and a correspondingly thicker insulating layer, where heat
transport is dominated by photons. The temperature dependence of the
metalization pressure indicates that, as WDs cool, the thickness of
the metallic layer increases at the expense of the insulating layer.
These results suggest that white dwarf stars may be 
older than previously thought.

\begin{acknowledgments}
  Financial support was provided by the Engineering and
  Physical Sciences Research Council (UK)\@. The calculations were
  performed on the Cambridge High Performance Computing Service
  facility.
\end{acknowledgments}

\bibliography{./anharmonic}

\end{document}